\documentclass[preprintnumbers,superscriptaddress,amsmath,amssymb,12pt]{revtex4}
\usepackage{graphicx}
\newcommand{\s}{\ensuremath{\psi(t,r)}}
\newcommand{\n}{\ensuremath{\nu(t,r)}}
\newcommand{\T}{\ensuremath{\theta}}

\newcommand{\lf}{\ensuremath{l_{6}}}
\newcommand{\e}{equation$\;$} 
\newcommand{\M}{\ensuremath{{\cal M}}}
\newcommand{\Ma}{\ensuremath{ m_{0}}}
\newcommand{\Mb}{\ensuremath{ m_{1}}}

\newcommand{\del}{\mbox{$\partial$}}

\def\be{\begin{equation}}
\def\eq{\end{equation}}
\def\ee{\end{equation}}

\begin{document}
\title{Critical Collapse of Einstein Cluster}

\author{Ashutosh Mahajan}
\email{ashutosh@tifr.res.in}
\affiliation{Tata Institute of Fundamental Research,
Homi Bhabha Road, Mumbai 400 005, India}

\author{Tomohiro Harada}
\email{harada@rikkyo.ac.jp}
\affiliation{Department of Physics, Rikkyo University, Toshima,
Tokyo 171-8501, Japan}

\author{Pankaj S. Joshi}
\email{psj@tifr.res.in}
\affiliation{Tata Institute of Fundamental Research,
Homi Bhabha Road, Mumbai 400 005, India}

\author{Ken-ichi Nakao}
\email{knakao@sci.osaka-cu.ac.jp}
\affiliation{Department of Physics, Osaka City University, Osaka 558-8585, Japan}

 \begin{abstract}
 We observe critical phenomena in spherically symmetric gravitational collapse of Einstein Cluster. We show analytically that the collapse evolution ends either in formation of a black hole or in dispersal depending on the values of initial parameters which characterize initial density and angular momentum of the collapsing cloud. Near the threshold of black hole formation, we obtain scaling relation for the mass of the black hole  and find the critical exponent value to be 3/2. We numerically confirm that there exist wide ranges of initial parameter values around the critical configuration for which the model remains shell-crossing free.

\end{abstract}
\maketitle

\section{Introduction}

Critical behavior in gravitational collapse was first found by Choptuik in  numerical simulations of spherically symmetric massless scalar field \cite{choptuik}. For marginally supercritical data, it was found that the black hole mass scales as $M \sim (p-p^{*})^{\gamma}$, where $p$ is a parameter of family of the initial data which characterizes strength of the configuration, and $p^{*}$ is the critical value. The critical exponent $\gamma$ has a value 0.37 for scalar field, and is universal in the sense that it is the same for all one parameter families of initial data.

Subsequently, the matter models were generalized. Critical behavior was found in the collapse of radiation fluid \cite{radiation}, perfect fluid with $p=k\rho$ \cite{perfect_fluid}, and in the collapse simulations of Yang Mills field with critical exponent 0.19 \cite{yang_mills} and also in axion-dilaton model from low energy effective string theory \cite{strings}. Although critical behavior has been found in numerical simulations of many matter models, very few analytical examples are available so far. Koike, Hara and Adachi gave explanation of the scaling of black hole mass and the universality by carrying out renormalization group studies \cite{rg}. These studies made it clear that the critical exponent for a given model can be obtained by finding the growth rate of the unique unstable mode of the self similar critical solution which is the fixed point of the renormalization group transformation. 
Subsequently, linear stability analysis became a standard technique to demonstrate critical behavior and calculate critical exponent in gravitational collapse models. On the other hand, it is not necessarily clear where and how the
linear order eigenvalue analysis gets applicable for initial data highly nonlinearly deviated from the critical solution, prior to numerical simulations.

 
Exact value of critical exponent has been calculated analytically in few models, which include a matter model with thin shell coupled with null fluid \cite{exact_exponent}, massless scalar field in four dimensions \cite{oshiro} and in (2+1) Ads Spacetime \cite{2+1}. We present here a model in which threshold of black hole formation can be seen and corresponding exponent can be calculated exactly by a different method \cite{tp}.

Here we consider a spherically symmetric collapsing model in which non-colliding particles move in such a way that the radial pressure vanishes and non-zero tangential pressure is present in the collapsing cloud. Such a static system was first introduced by Einstein \cite{einstein} which was later generalized to non-static case \cite{Bondi}. In recent years, the properties of this model have been studied extensively \cite{hin,jhingan}.

We observe that the model shows criticality near the threshold of black hole formation. We work in the comoving coordinates and investigate evolutions of individual shells. It is  analytically shown that the collapsing cloud either forms a black hole or disperses depending on the values of initial parameters which characterize the initial density and
tangential pressure of the collapsing cloud. We derive scaling relation for the mass of the black hole near its threshold.

 This model has a limitation that the radial stress is vanishing and the Misner-Sharp mass for the collapsing cloud is time independent. However, this simplifies the Einstein equations and make the model tractable. 
The model considered here is different from the models considered for the criticality so far in the following aspects. Critical exponent is obtained without invoking self similarity. We have found the critical self similar solution for this system and it will be discussed elsewhere\cite{ss_ecluster}. The no-shell crossing conditions make the initial data restrictive, and only certain ranges of initial data is allowed in the comoving coordinate system. The model shows critical behavior for two parameters as long as we assume regularity at the initial epoch and the smoothness of the initial parameters, and thus, in the space of initial data sets, the requirement of black hole formation within this model already restricts the space of possible sets to be of codimension one, in contrast with the standard critical behavior where the shape of the initial data have very large variety. Despite all these limitations, the model remains interesting as one can see transparently how an infinitesimal mass black hole formation can take place when we fine tune the initial data.

The outline of the paper is as follows. In Section II, we discuss collapse equations and regularity conditions and in Section III, the Einstein cluster model is discussed. In Section IV, critical behavior is investigated and it is demonstrated how a given sets of initial value parameters decide the final outcome. In Section V scaling law for the black hole mass is derived. Shell crossing is discussed in Section VI and conclusions are outlined in Section VII.

\section{Einstein Equations, Regularity and Energy conditions}

We use the polar coordinates $(t,r,\theta,\phi)$ to write the 
spherically symmetric metric as
\begin{equation}
ds^2= -e^{2\n}dt^2 + e^{2\s}dr^2 + R^2(t,r)d\Omega^2,
\label{metric}
\end{equation}
where $d\Omega^2$ is the line element on two-sphere. We also
take the energy-momentum tensor to be diagonal for 
the collapsing {\it Type I} matter field 
(that is, the frame is a comoving coordinate 
system) which is given by
\begin{equation}
T^t_t=-\rho;\; T^r_r=p_r;\; T^\T_\T=T^\phi_\phi=p_\T.
\label{setensor}
\end{equation}
This is a fairly general class of matter fields, which includes
various known physical forms of matter
\cite{he}.
The quantities $\rho$, $p_r$ and $p_\T$ are density, radial 
pressure and tangential pressure respectively. We take the matter 
field to satisfy the {\it weak energy condition}, that is, the energy 
density as measured by any local observer be non-negative, and so 
for any timelike vector $V^i$ we have
\begin{equation}
T_{ik}V^iV^k\ge0.
\end{equation}
This amounts to
\begin{equation}
\rho\ge0;\; \rho+p_r\ge0;\; \rho+p_\T\ge0.
\end{equation}
The dynamical evolution of the system is determined by the 
Einstein equations, and for the metric (\ref{metric}) in the  units
$8 \pi G=c=1$, these are given as
\be
\rho = \frac{F'}{R^2R'}, \; \; \; \;
  p_{r}=-\frac{\dot{F}}{R^2 \dot{R}},
\label{t6}
\eq
\be
\nu'(\rho+ p_{r})=2(p_{\theta}-p_{r})\frac{R'}{R}-p_{r}',
\label{t7}
\eq
\be
-2 \dot{R'}+R'\frac{\dot{G}}{G}+\dot{R}\frac{H'}{H}=0,
\label{t8}
\eq
\be
G-H=1 - \frac{F}{R},
\label{t9}
\eq
\\
where $(\,\dot{}\,)$ and $(')$ represent partial derivative with respect 
to $t$ and $r$ respectively and
\be
G(t,r)=e^{-2\psi}(R')^2, \; \; H(t,r)=e^{-2\nu}\dot{R}^2.
\eq       
The quantity $F(t,r)/2$ is the Misner-Sharp mass for the collapsing 
cloud, which gives total mass within a shell of  comoving radius 
$r$ at time $t$ \cite{misner}. In order to preserve the regularity at the 
initial epoch, 
$F(t_i,0)=0$, that is, the mass function should vanish at 
the center of the cloud. It can be seen from the equation ($\ref{t6}$) 
that density of the matter blows up when $R=0$ or $R'=0$. Here the 
case $R'=0$ corresponds to the {\it 
shell-crossing} singularities.

Now let us write physical radius as
\begin{equation}
R(t,r)=rv(t,r).
\label{R}
\end{equation}
Using the scaling independence of the coordinate $r$ and initial 
collapse condition, we write
\begin{eqnarray}
v(t_i,r)=1; \, & v(t_s(r),r)=0; \, & \dot{v}(t_i,r)<0,
\label{v}
\end{eqnarray}
where $t_i$ and $t_s$ stand for the initial and the singular   
epochs respectively.  The condition $\dot{v}(t_i,r)<0$ signifies
that we are dealing with initially collapsing shells.
We scale the radial coordinate $r$ in such a way that at 
the initial epoch $R=r$, and at the singularity, $R=0$. 
The advantage of the introduction of this new variable
$v$ is that the regular center at $r=0$ (where we also
have $R=0$) is now distinguished
from the genuine singularity at $R=0$. We now have
$v=1$ at the initial epoch, and $v=0$ at the singular
epoch $R=0$, but at all other epochs in-between $v$ has a 
non-zero finite value for all values of $r$.

\section{Non-static Einstein Cluster Model}

The spherically symmetric collapse models, where the radial 
pressure is taken to be vanishing and the tangential pressure could be 
non-zero have been studied in quite some detail over past many years 
\cite{hin,jhingan,tan}.

The Einstein cluster is an example of such a cloud where 
tangential stresses are present. This is a spherically symmetric cluster 
of rotating particles where the motion of the particles is sustained 
by an angular momentum which has an average effect of creating a 
non-zero tangential stress within the cloud. Neighboring shell particles 
are counter-rotating such that spherical symmetry is preserved.

We consider such a non-static cluster of gravitating particles in four 
dimensions. For the non-static Einstein cluster models the equation of state
 is given by \cite{Bondi}
\be
p_{\theta}=\frac{1}{2}\left(\frac{L^2}{R^2+L^2}\right)\rho,
\label{cluster1}
\ee
where $L(r)$ is a function of the radial coordinate $r$ only
and is known as {\it specific angular momentum}.Vanishing radial pressure implies that the Misner-Sharp mass is time independent. Regularity of the initial density at the center requires
\be
F(r)=r^3 \M(r),
\eq 
where $\M(r)$ is a smooth function. It is clear that as $v\rightarrow 0$, 
$\rho\rightarrow\infty$. Thus the density blows up at the singularity $R=0$ 
which will be a curvature singularity as expected. Let us now define a suitably differentiable function $A(r,v)$ 
in the following manner
\begin{equation}
\nu'(r,v)=A(r,v)_{,v}\,R'.
\label{eq:A}
\end{equation}
Then from equation (\ref{t7}), we have the equation of state given as
\be
p_{\theta}=\frac{1}{2}A_{,v}R\, \rho.
\label{eqstate}
\ee
Now using the equation 
(\ref{eq:A}), we can integrate (\ref{t8}) to get
\begin{equation}
G=b(r)e^{2rA}.
\label{eq:G}
\end{equation}
Here $b(r)$ is another arbitrary function of the comoving coordinate
$r$. Following a comparison
with dust collapse models we can write
\begin{equation}
b(r)=1+r^2b_0(r),
\label{eq:veldist}
\end{equation}
where $b_0(r)$ is the energy distribution function for
the collapsing shells.
Finally, using equations (\ref{eq:A}), (\ref{eq:G}) and 
(\ref{eq:veldist}) in
(\ref{t9}) we have
\be
R{\dot{R}}^{2} e^{-2\nu}= {(1+r^2b_0)Re^{2rA}-R+
r^{3}\M}.
\label{collapse1}
\ee
A comparison of equation (\ref{cluster1}) with (\ref{eqstate}) gives
\be
A_{,v}=\frac{L^2}{R(R^2+L^2)}.
\label{A2}
\ee
We can integrate above equation to get
\be
e^{2rA}=\frac{R^2}{R^2+L^2}.
\label{A1}
\ee
Considering initial density, pressure and energy profiles
to be smooth would ensure $L(r)$ also to be smooth.

We write the initial profiles in the form 
\be
\M(r)=m_{0}+m_{2}r^2+\cdots,
\label{mass1}
\eq 
\be
L^2(r)=l_{4}r^4+ l_{6}r^6 + \cdots, 
\label{ls}
\ee
\be
b_{0}(r)=b_{00}+b_{02}r^2+\cdots.
\label{velocity}
\eq 
The initial density profile, initial specific angular momentum 
profile and velocity profile of the cloud are chosen fully in equations 
(\ref{mass1}), (\ref{ls}) and (\ref{velocity}) respectively. Regularity 
at the initial epoch requires leading order term of $L^2$ to go as $r^4$. 
Now we need to evolve this initial data according to equation 
(\ref{collapse1}).

\section{Critical Phenomena}

Initially, at the onset of gravitational collapse all the shells 
have the scale factor $v(t_i,r)$ value as unity and $\dot{v}(t_i,r)<0$,
which implies an initially collapsing cloud. Bounce of a shell is 
indicated by the change in the sign of $\dot{v}$. 
Evolution of a particular shell may be deduced from 
\e (\ref{collapse1}). Rewriting (\ref{collapse1}) in terms of 
the scale factor we have

\be
 V(r,v)=-{e^{-2\nu}}{\left(v^{2}+\frac{L^{2}}{r^{2}}\right)}v\dot{v}^{2}.
\label{pot1}
\eq
We call $V(r,v)$ as the {\it effective potential} for a shell.
It can be expressed in terms of initial profile functions of the system 
as follows

\be
V(r,v)= -\left(b_{0}\, v^{3} + \M \,v^{2} -\frac{L^{2}}{r^{4}} \, 
v+ \frac{\M L^{2}}{r^{2}}\right). 
\label{gendyn}
\eq

The allowed regions of motion correspond to ${V}(r,v)\leq0$, as 
$v \dot{v}^{2}$ is non-negative.
The first factor in equation (\ref{pot1}) is 
always positive, because it is the $g^{00}$ term of the metric 
tensor and the quantity in parenthesis in the same equation is also always positive, hence location of the turning points (where $\dot{v}$ changes its sign) are not decided by these two terms. The main features of the evolution of a shell 
basically derive from the cubic polynomial in equation (\ref{gendyn}).

The dynamics of the shells may be studied by finding the concerned  
turning points. If we start from an initially collapsing state ($\dot{v}<0$), 
we will have rebounce for a shell if we get  $\dot{v}=0$, before 
the shell has become singular. This can happen when ${V}(r,v) = 0$ (see 
for example Fig.1). Hence, to study the various evolutions for a particular 
shell we must analyze the roots of the equation ${V}(r,v) = 0$, keeping 
the value $r$ to be fixed. The method employed here is similar 
to another class of tangential pressure collapse models where the 
metric function $g_{00}$ is taken to be a function of the physical 
radius $R$ 
\cite {tp}.

Out of three roots of the cubic polynomial only positive 
real roots correspond to physical cases. We take $b_{0}$ to be positive.
The region between the unique positive roots is forbidden since in those
regions $\dot{v}^{2}<0$. For a particular shell to bounce, it must therefore 
lie, during initial epoch ($v=1$), in a region to 
the right of the second positive root.
We can see that three different types of evolution of a particular shell are 
possible:\\

i) If the effective potential for a shell has two distinct positive roots in the range $[0,1]$, then the shell bounces off.




ii) If $V(r,v)< 0$ in the whole range $[0,1]$, then the shell 
will reach the singularity at $v=0$.

iii) When the potential has a double root in $[0,1]$, then it indicates that 
the shell is in critical collapse condition.\\

Let us assume that we are working in initial data space for which there are no
 shell-crossings. We will discuss the issue of shell-crossings in the matter cloud in \S6. If a particular shell with comoving coordinate $r_a$
 bounces then all the shells with coordinate $r>r_a$ must also bounce.
 This implies that to investigate the situation when the entire cloud
 is just about to disperse off, it is 
sufficient to study the dispersal of the shells near the center. 
Therefore, to find the threshold of black hole formation and to get 
the scaling relation for the mass of the black hole, we need to 
analyze the model only near the center. We have neglected here higher 
order terms in $r$ in the expansion, since we want to consider 
only the evolution of the shells near $r=0$.

To analyze the dynamics in this model in detail, we study one 
particular configuration of the initial data, in which $\Ma$, $m_{2}$ 
are kept fixed, $b_{00}$ is taken positive, $l_{4}$ is set 
to zero and $\lf$ is allowed to vary. Non-zero value of $l_{4}$ will 
make the central shell and hence the cloud to always bounce and 
black hole will not be formed. Close to the center, we neglect higher 
order terms in the expansion, in which case effective potential 
can be written as 

\be
V \approx -b_{0}(r)\, v^{3} - (m_{0}+m_{2}r^2) \,v^{2} +l_{6}r^2 \, 
v - (m_{0}+m_{2}r^2) l_{6}r^4, 
\label{dyn}
\eq
where $b_{0}(r)= b_{00}+b_{02}r^2$. We use an arbitrary unit since there is no length scale in the original system of equations.
 For a cubic equation $ax^3+bx^2+cx+d=0$, a double root occurs when discriminant $\Delta$ for the polynomial vanishes, where 
\be
\Delta= 18abcd-4b^{3}d+b^{2}c^{2}-4ac^{3}-27a^{2}d^{2}.
\label{delta}
\eq
For above cubic equation this takes the form
\be
\Delta = r^4 \Delta_{1}(r,p_i^x),
\label{delta1}
\eq
where $p_i^x$ are the initial parameters. Now we study collapse 
evolutions for three different categories of the initial data $p_i^s, p_i^c$ 
and $p_i^b$.\\
\vspace{.5cm}

{\it \bf A) Supercritical Evolution} \\
 
Fig.1 shows the effective potential for the central shell 
$(r=0)$ and for three outer shells $(r=0.10,0.113,0.15)$ for a small value of $l_6$. Potential for the central shell has always a double root ($r=0, \Delta_1(0,p^s_i)\neq0$). If we go slightly away from the center, potential for a shell goes negative for the entire range of $v$ which allows the shell to reach the singularity at $v=0$.

If we increase $r$ further, the potential maxima starts going 
up, and there is a value of $r$ at which the effective potential 
again has a double root. We call this radius as the critical radius $r_{c}$ of the collapsing cloud for the chosen set of the initial numbers. At this value the discriminant of the cubic vanishes ($\Delta_1(r_c,p^s_i)=0$).

Apparent horizon is given by 
\be
F=R,
\ee 
which near the center in $(v,r)$ plane can be written as $v_{ah}(r)=r^2 m_0$,
where at $v=v_{ah}$ a shell becomes trapped.
It can be seen from the figure that if the potential remains 
negative in the range [0,1], the value of $v$ decreases and finally goes below $v_{ah}$ which indicates the trapping of the shell. Now as the shells below $r_{c}$ go inside the apparent horizon, a black hole eventually forms. These shells finally reach the singularity and contribute for the mass of the black hole formed, but the shells with comoving coordinate more than critical radius bounce off. 
As the black hole forms for these values of $\lf$, this represents supercritical region.\\

\begin{figure}[h!!]
 \centerline{\includegraphics[width=7.0cm,angle=-90]{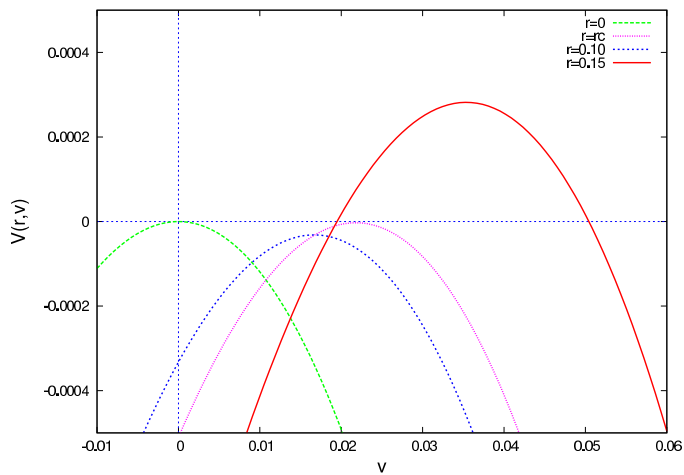}}
\caption{\small{Figure represents effective potential for different comoving radius in supercritical configuration. Here $l_{6}=3.5$, $b_{0}=4.0$, $m_{0}=1.15$ and $m_{2}=-20.$}}
\end{figure}

{\it \bf B) Critical Evolution} \\

As we increase $\lf$ further, we reach a value at which the 
potential for the central shell has double root ($r=0,\Delta_1(0,p^c_i)= 0$) and all other shells have two distinct real positive roots. We call this value as the critical value $l_{6c}$ of the initial parameter $\lf$ for the other fixed parameters. This situation is shown in Fig.2. The outer shells have positive potential and forbidden region, therefore, all those shells will bounce back. This configuration gives 
the critical solution of the system. The critical point is the boundary point between the dispersal and the situation when the collapse to form black hole just begins.\\

\begin{figure}[h!!]
\centerline{\includegraphics[width=7.0cm,angle=-90]{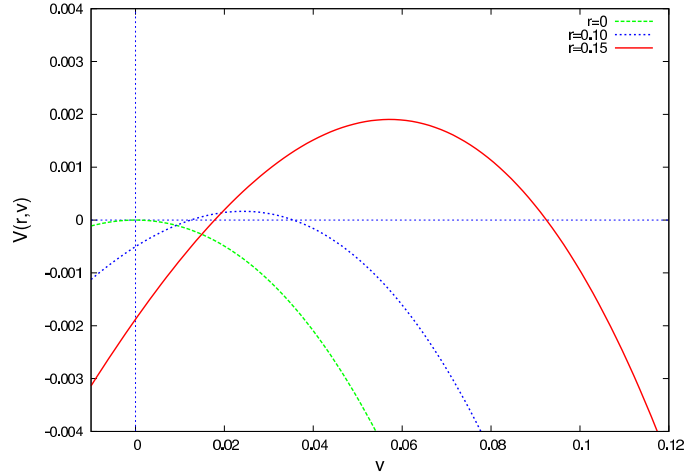}}
\caption{\small{Figure represents effective potential for different comoving 
radius in critical configuration. Here $l_{6}=5.3$, $b_{0}=4.0$, $m_{0}=1.15$ 
and $m_{2}=-20.$}}
\end{figure}

{\it \bf C) Subcritical Evolution} \\

If we increase $l_6$ further to $l_{6c}$ potential goes even 
more positive. There is a complete bounce of the collapsing shells. Hence this initial configuration generates subcritical evolution. However it should be noted that only the central shell reaches the singularity ($\Delta_1(0,p^b_i)\neq0$) and all other shells bounce. This massless central singularity is a timelike naked singularity which can be seen as follows.
For the central shell $e^{\nu(0,v)}v\dot{v}^2=b_{00}v+m_0$, and as $m_0$ and $b_{00}$ are positive, central shell reach singularity at $v=0$ while all outer shells bounce off at value of $v$ which is larger than $v_{ah}$ for those shells and are not trapped. Thus, central singularity forms but trapped surfaces do not form in the cloud, making the singularity globally visible \cite{hin}. 

\begin{figure}[h!!]
 \centerline{\includegraphics[width=7.0cm,angle=-90]{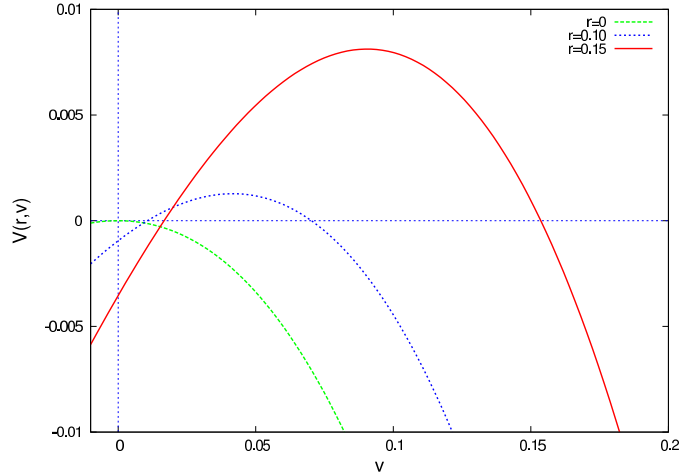}}
\caption{\small{Figure represents effective potential for different comoving 
radius in subcritical configuration. It can be seen that outer shells bounce. 
Here $l_{6}=10$, $b_{0}=4.0$, $m_{0}=1.15$ and $m_{2}=-20.$}}
\end{figure}

\section{Mass Scaling Law}

To derive the scaling law for black hole masses, it is the key to
estimate the critical shell radius, which separates the completely
collapsing cloud from the surrounding dispersive cloud. The expression for the critical radius in terms of the initial 
parameters can be obtained in the supercritical region from the condition 
that at critical radius, effective potential has two equal roots. The double root condition for polynomial in (\ref{dyn}) together with the consideration of shells only near the center gives

\be
r_{c}^2 \approx \frac{4m_{0}^4-m_{0}^{2} \,l_{6}}
{-16m_{0}^{3}m_{2}-18\,b_{00}m_{0}^{2}l_{6}+2\,m_{0}m_{2}l_{6}+4\,b_{00}l_{6}^2}.
\eq
At $\lf=l_{6c}=4m_{0}^2$, the critical radius vanishes. 
It can be seen in Fig.4 that as we increase $l_6$ in the supercritical 
region, the critical radius monotonically decreases from a positive 
value and becomes zero at $l_{6c}$. If we increase $l_6$ further all the non-central shells in the cloud bounce. Therefore, we can write 

\be
r_{c}^2 \approx k \, {|\lf-l_{6c}|},
\eq
where $k$ is a constant. In the tangential pressure model, the 
Misner-Sharp mass depends only on $r$. Mass which collapsed to form 
singularity from the regular initial profile is given by

\begin{figure}[h!]
 \centerline{\includegraphics[width=7.0cm,angle=-90]{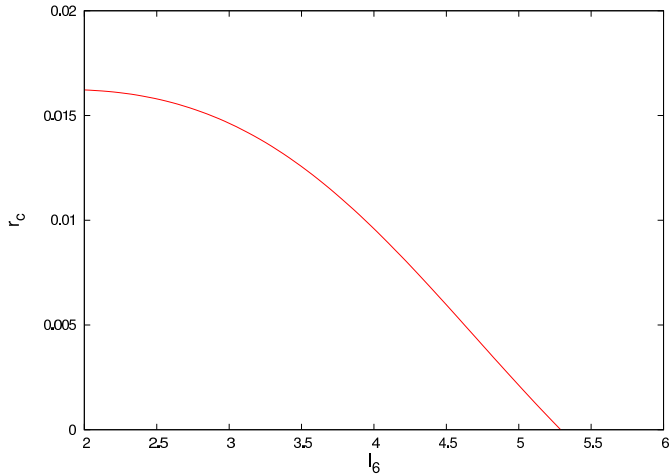}}
\caption{\small{Critical radius in the supercritical region. The curve 
is the locus of the point on which $\Delta_1$ vanishes.}}
\end{figure}
\be
F(r_{c})/2 \approx \frac{1}{2}r_{c}^{3}(m_{0}+m_{2}r_{c}^{2}).
\eq
As we are analyzing the system near the center and also near the threshold of black hole formation, critical radius $r_c\rightarrow 0$, therefore the mass of such black hole can be written as
\be
M_{BH} \approx c_{m}|\lf-l_{6c}|^{\frac{3}{2}},
\eq
where $c_{m}$ is a constant near the threshold.

We can fix any of the parameters  $\Ma$ and $\lf$ and vary the 
other one to obtain the expression for critical radius. Following the 
same procedure as depicted earlier, it can be easily seen that 
for $\Ma$, the same scaling relation and critical exponent exists near 
the threshold of the black hole formation. Therefore, in general for the 
parameter $\eta$, we write  

\be
M_{BH} \approx c_{\eta}|\eta-\eta_{c}|^{\frac{3}{2}}.
\eq 

\section{Shell Crossings}

Above analysis is based on the condition that for the chosen initial data near the critical values, there are no shell crossings during the evolution of the collapsing cloud, which implies that if a shell bounces then all the shells with a larger value of comoving radius will also bounce. It is important to check that there are indeed no shell crossings taking place for a given initial data. We know that if $R'(r,t)>0$ there are no shell crossings, therefore, we need to obtain $R'$ for a given set of initial conditions. Equations (\ref{eq:A}) and (\ref{collapse1}) are two coupled equations for $\nu$ and $R(r,t)$. They can not be solved explicitly analytically in the coordinate system used and numerically also solving them is not easy.

 However, there is a way out, Gair has given a coordinate transformation which eliminates the function $\nu$ in the evolution equation, using which one can obtain $R'$ at constant time $t$ in the following manner \cite{gair}.
He has used a time coordinate $\tau$ which he calls proper time experienced by a dust particle
\be
\left(\frac{\del\tau}{\del t}\right)_r=\frac{e^{\nu}}{\sqrt{1+\frac{L^2}{R^2}}}.
\label{tau}
\eq
After changing the coordinates, we get from equation (\ref{collapse1})
\be
\left(\frac{\del R}{\del \tau}\right)^2_r=r^2b_{0}+\frac{F}{R}\left(1+\frac{L^2}{R^2}\right)-\frac{L^2}{R^2}.
\label{evolution}
\ee
 We can choose a common origin of time for all shells, i.e. at some time, $t=0$, we set $\tau=0$ for all r (see section 3 in \cite{gair}). This gives us the initial conditions to integrate above equation numerically.
 We note that
\be
e^{-\nu}\sqrt{1+\frac{L^2}{R^2}}\left(\frac{\del}{\del t}\right)_r=\left(\frac{\del}{\del \tau}\right)_r,
\label{partial1}
\ee
\be\left(\frac{\del R}{\del r}\right)_t=
\left(\frac{\del R}{\del r}\right)_{\tau}+\left(\frac{\del \tau}{\del r}\right)_t \left(\frac{\del R}{\del \tau}\right)_r.
\label{chain}
\ee

Now, to get the desired quantity what remains is to know $({\del \tau}/{\del r})_t $. To obtain it, we first differentiate equation(\ref{tau}) with respect to r, then change the order of partial derivatives and use equation (\ref{partial1}), obtaining a differential equation for  $({\del \tau}/{\del r})_t$ which is as the following

\be
\left(\frac{\del}{\del \tau}\right)_{r}\left(\frac{\del \tau}{\del r}\right)_t
=\left[ \xi(r,\tau) \left(\frac{\del R}{\del \tau}\right)_r -\left(\frac{\del \psi}{\del \tau}\right)_r
 \right]\left(\frac{\del \tau}{\del r}\right)_t+\left[\xi(r,\tau) \left(\frac{\del R}{\del r}\right)_{\tau} -\left(\frac{\del \psi}{\del r}\right)_{\tau}
 \right],
\label{diffeqn}
\ee
where $\xi$ is $\frac{L^2}{R(R^2+L^2)}$ and $\psi$ is $\frac{1}{2}\left(1+\frac{L^2}{R^2}\right)$. The initial condition for the above equation is provided by the fact that we have set $\tau=0$ at $t=0$ for all the shells, which implies $(\del \tau/ \del r)_t=0$ at $\tau=0$. We integrate (\ref{evolution}) first and then substitute the values of $(\del R/\del r)_{\tau}$ and $(\del R/\del \tau)_r$ in equation (\ref{diffeqn}), after integrating which equation (\ref{chain}) gives $(\del R/\del r)_t$.\\
\begin{figure}[h!!]
\centerline{\includegraphics[width=7.0cm,angle=-90]{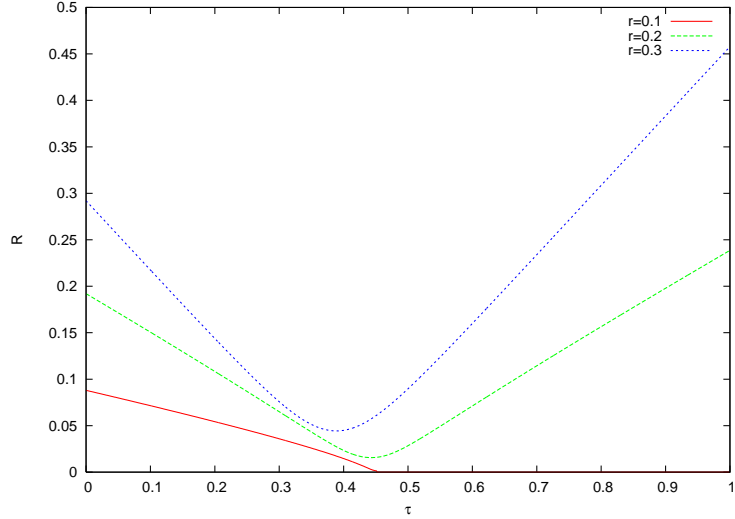}}
\caption{\small{Physical radius of different shells in supercritical evolution, where initial parameter values are $\Ma=1.15,b_{0}=1.95, b_{2}=50$, $\Mb=5.0$ and $L=5.0$. It can be seen that shells with comoving radius less than the critical radius reach singularity while those with larger than critical radius bounce.}}
\end{figure}
\begin{figure}[h!!]
\centerline{\includegraphics[width=7.0cm,angle=-90]{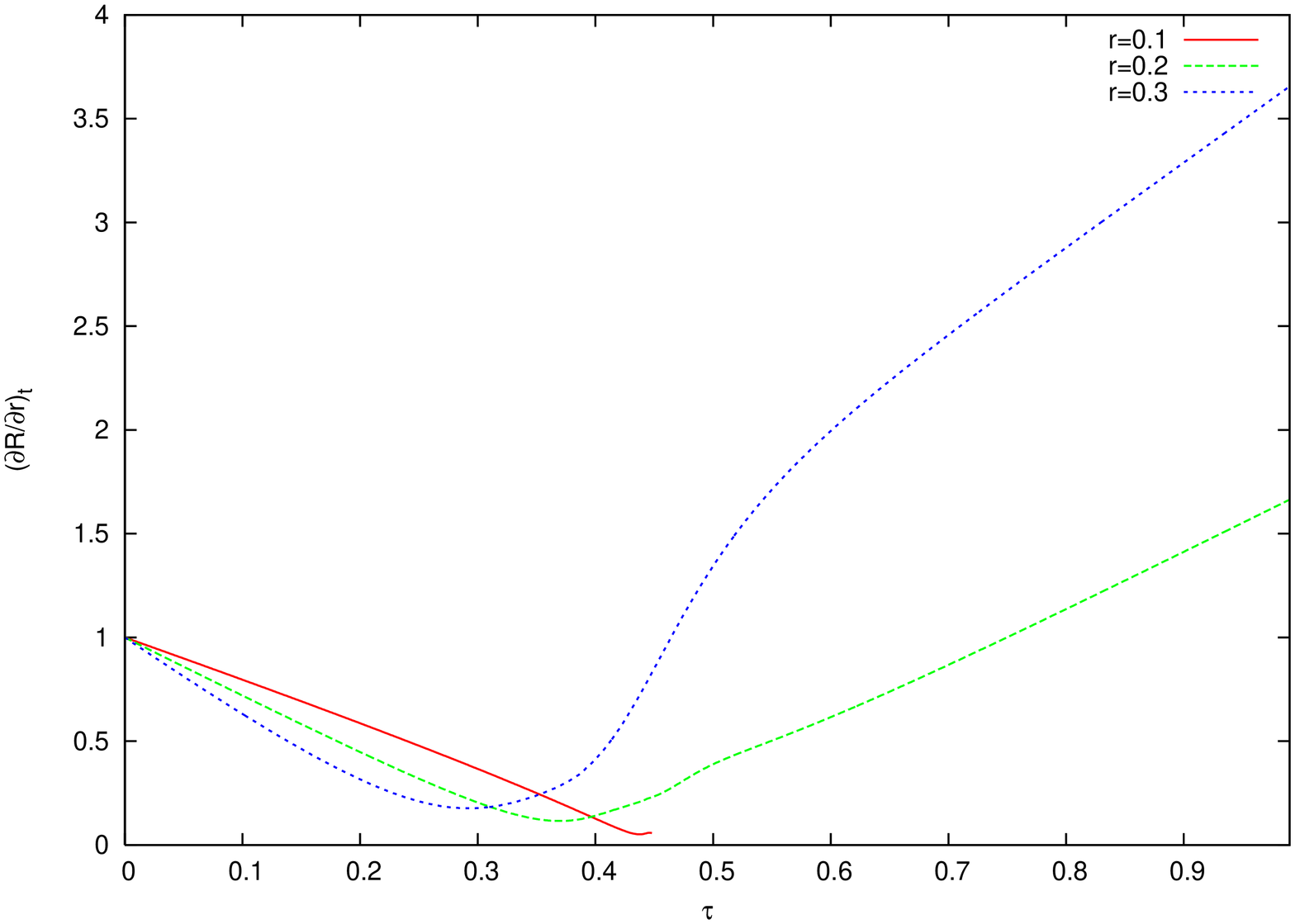}}
\caption{$(\del R/\del r)_t$ in supercritical evolution, where the initial parameter values are $\Ma=1.15,b_{0}=1.95,b_{2}=50$, $\Mb=5.0$ and $L=5.0$. It can be seen that the value of $R'$ at constant time $t$ remains positive.}
\end{figure}

\begin{figure}[h!!]
\centerline{\includegraphics[width=7.0cm,angle=-90]{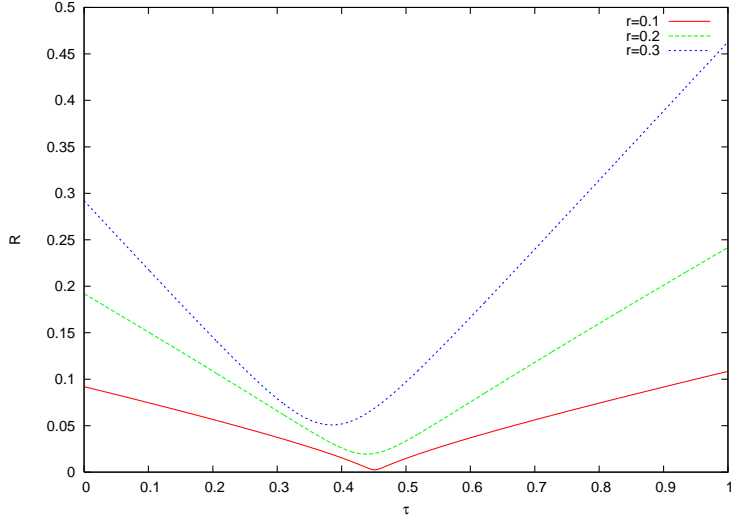}}
\caption{\small{Physical radius of different shells in subcritical evolution, where initial parameter values are $\Ma=1.15,b_{0}=1.95,b_{2}=50$, $\Mb=5.0$ and $L=5.7$. It can be seen that all the shells bounce.}}
\end{figure}
\begin{figure}[h!!]
\centerline{\includegraphics[width=7.0cm,angle=-90]{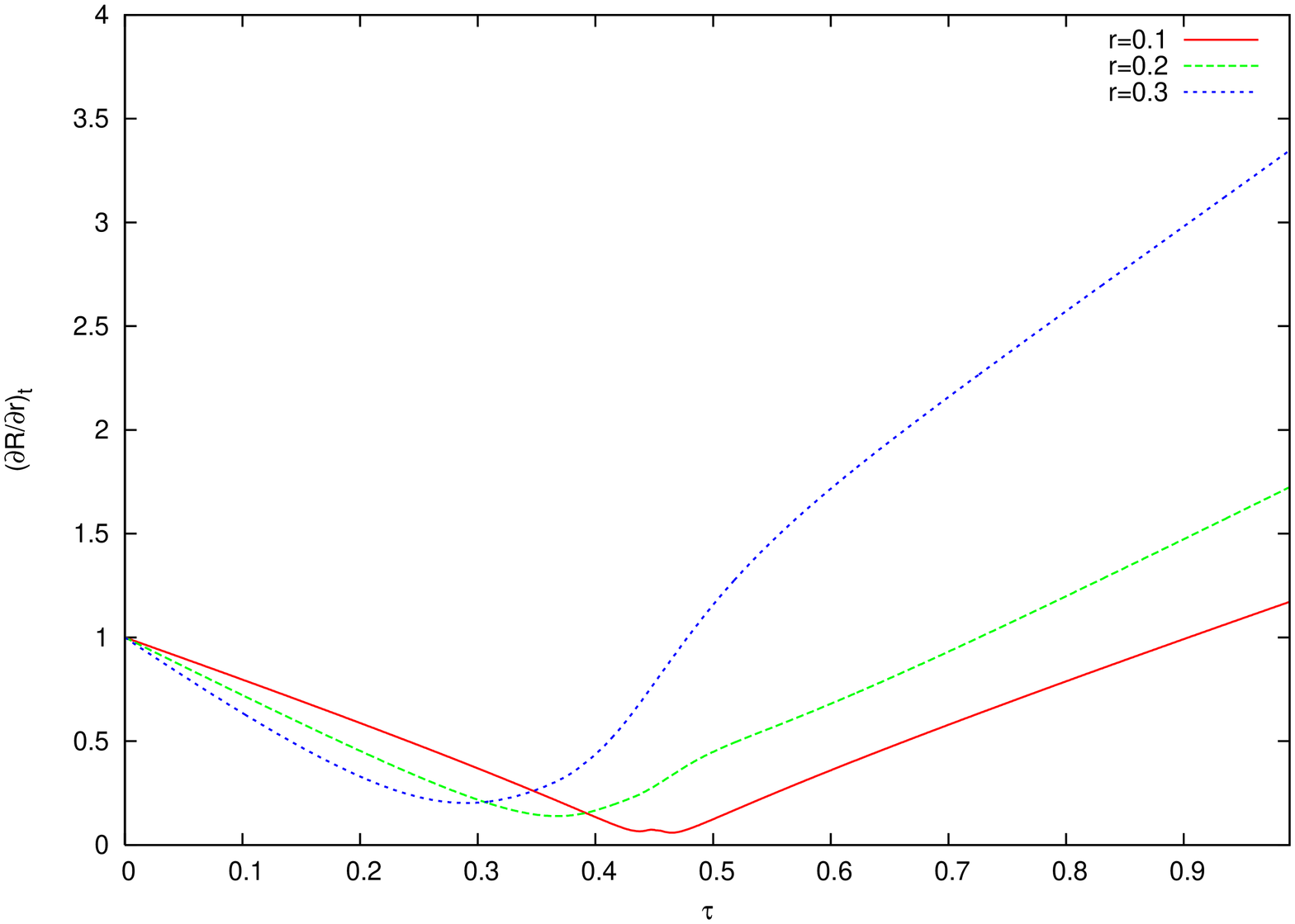}}
\caption{\small{$(\del R/\del r)_t$ in subcritical evolution, where the initial parameter values are $\Ma=1.15,b_{0}=1.95,b_{2}=50$, $\Mb=5.0$ and $L=5.7$. It can be seen that the value of $R'$ at constant time $t$ remains positive.}}
\end{figure}

 We find that for the chosen initial data in the vicinity of critical situation, there are no shell crossings. Physical radius $R$ and $(\del R/\del r)_t$ for the subcritical and supercritical case are plotted in the following figures.
 We do not give here a general criteria or condition on the initial parameter functions which will ensure no shell crossings. However, one can choose the initial data which gives no shell crossings and one can see that the critical solution lies in this allowed range of parameters.

\section{Conclusions}

In this paper, we have presented a simple analytic model of 
gravitational dynamics which is physically motivated in the sense 
that it is a special realization of the Einstein-Vlasov (collisionless 
particles) system. The system satisfies standard requirements of physical 
reasonableness, namely, it has an equation of state, and satisfies 
energy and regularity conditions.

We have shown that both black hole and dispersal are possible outcomes in the gravitational collapse of this model. The mass of the black hole near the threshold of its formation shows power law behavior with critical exponent 3/2. Very small mass black holes could be formed if the initial data is fine tuned. 
The model shows "universal" behavior with respect to the parameters 
which fix the initial density and specific angular momentum of the cloud. As shown in the plots, for a generic class of the regular initial
data involving central singularity formation, shell crossing does not occur.
The effective potential method for calculating the critical exponent 
applies to this case as well as to an another class of tangential pressure models \cite{tp}. This suggests that the method applies at least for all 
mass-conserving systems. However, the system considered here has a 
limitation and bit different from the models considered so far for 
critical behavior in one respect that it has no radial stress 
which enables shells to interact directly with each other. However,
the critical behavior seen in general relativity appears to be 
a generic phenomenon and it is expected that only the radial 
stress is not essential for the phenomena to occur. It will be very 
interesting to explore whether the same method holds for 
studying critical behavior in the model with radial pressure as well.

\section*{Acknowledgments}

We would like to thank Tatsuhiko Koike, Susan Scott and Naresh Dadhich for helpful discussions.TH was partly supported by the Grant-in-Aid for Scientific
Research on Priority Areas, 14047212 and for Young Scientists
(B), 18740144 from the Ministry of Education, Culture, Sports,
Science and Technology (MEXT) of Japan.


\end{document}